# Phoenix - The Arabic Object-Oriented Programming Language

Youssef Bassil

*LACSC – Lebanese Association for Computational Sciences*
*Registered under No. 957, 2011, Beirut, Lebanon*

*Abstract*

*A computer program is a set of electronic instructions executed from within the computer's memory by the computer's central processing unit. Its purpose is to control the functionalities of the computer allowing it to perform various tasks. Basically, a computer program is written by humans using a programming language. A programming language is the set of grammatical rules and vocabulary that governs the correct writing of a computer program. In practice, the majority of the existing programming languages are written in English-speaking countries and thus they all use the English language to express their syntax and vocabulary. However, many other programming languages were written in non-English languages, for instance, the Chinese BASIC, the Chinese Python, the Russian Rapira, and the Arabic Loughaty. This paper discusses the design and implementation of a new programming language, called Phoenix. It is a General-Purpose, High-Level, Imperative, Object-Oriented, and Compiled Arabic programming language that uses the Arabic language as syntax and vocabulary. The core of Phoenix is a compiler system made up of six components, they are the Preprocessor, the scanner, the parser, the semantic analyzer, the code generator, and the linker. The experiments conducted have illustrated the several powerful features of the Phoenix language including functions, while-loop, and arithmetic operations. As future work, more advanced features are to be developed including inheritance, polymorphism, file processing, graphical user interface, and networking.*

**Keywords -** *Arabic Programming, Compiler Design, Object-Oriented, Programming Languages.*

## I. INTRODUCTION

When computers were first designed, they were all hardwired, in that they were limited to perform predefined functionalities without being able to be controlled or manipulated by software. After several decades, programmable computers were finally invented [1]. In early days of programmable computers, programming was not done through software as it is being done today; rather, it was done by configuring a combination of plugs, wires, and switches. For instance, in order to perform an addition operation, a cable has to be manually connected from a central hub to the adder unit [2]. As a result, controlling and setting up new tasks were very challenging and time-consuming. As the years have gone by, a brilliant scientist came up with a genius idea in the late 1940s, he thought that he could automate the programming tasks in a computer by using encoded instructions stored in computer's memory and executed sequentially to perform certain operations. John von Neumann called his breakthrough "Stored-Program" [3]. The stored-program concept means that instructions that make up the software are stored electronically in binary format in the computer's memory, rather than being manually configured by humans using wires and knobs from outside the computer. The idea was further developed to incorporate both data and instructions in the same memory, a model that is known as the Von Neumann Architecture [4].

Fundamentally, a computer program as we know it today is a set of electronic instructions executed from within the computer's memory by the computer central processing unit CPU. The purpose of a computer program is to control the functionalities of the computer allowing it to perform miscellaneous tasks including mathematical computations, scientific operations, accounting, data management, gaming, text editing, audio, video, and image archiving, and Internet. Basically, a computer program is written by a human using a programming language. A programming language is the set of grammatical rules and vocabulary that governs the correct writing and structure of a computer program or code [5]. A trivial property of a programming language is the human language it uses to express its syntax and vocabulary. For instance, the programming language C uses the English language as a means to write code. Another example is the Chinese BASIC which uses the Chinese language to write computer programs.

This paper discusses the design and implementation of a new programming language, called Phoenix. It is a General-Purpose, High-Level, Imperative, Object-Oriented, and Compiled Arabic programming language that uses the Arabic language as a means to write computer programs.

## II. EXISTING NON-ENGLISH PROGRAMMING LANGUAGES

Non-English programming languages are

programming languages that do not use the English vocabulary to write programming statements. Over the past decades, several non-English programming languages were developed with the purpose to appeal to the local audience especially students and non-English speakers. For instance, ALGOL 68 was extended to support several natural languages other than English such as Russian, German, French, and Japanese [6]. In the early 1970s, Chinese programming languages were introduced to make learning programming easier for Chinese programmers. Some of these languages include Chinese BASIC [7], Easy Programming Language (EPL) [8], and ChinesePython [9]. In French, there also exist couple of programming languages whose syntax is written in the French language. Linotte [10] for instance is an interpreted high-level language targeted to French-speaking children to easily learn programming in their native language. Likewise, LSE (Language Symbolique d'Enseignement) is a French programming language similar to BASIC exhibiting some advanced features such as functions, conditional statements, and local variables [11]. Furthermore, hundreds of programming languages exist using international languages though none of them has gone mainstream, such as Hindi Programming, a programming language using Hindi syntax [12], Mind [13], a Japanese programming language, Latino [14], a language based on Spanish syntax and vocabulary, Rapira [15], a Russian-based programming language mainly intended for educational usage in schools, and VisuAlg [16], a Portuguese-based programming language similar to Pascal, designed for educational purposes.

As concerning Arabic-based programming languages, several were presented. One of the earliest attempts to develop an Arabic programming language was by Al Alamiah company, a Kuwaiti leading company in Arabic language technologies who developed the Arabic Sakhr Basic in 1987 [17]. Sakhr Basic is an Arabized version of the BASIC language with keywords and expressions written using the native Arabic language. It targeted the Arabic version of MSX home computers originally conceived by Microsoft. ARLOGO [18] is another Arabic programming language intended for educational purposes and is based on UCB Logo language. ARLOGO is open-source and currently available only for Microsoft Windows. ARABLAN [19] is yet another Arabic programming language designed in 1995 and planned for use in teaching programming for school children in the Arab countries. Al-Risalah [20] is an Arabic object-oriented programming language providing the basic mechanisms of object-orientation including classes, objects, and composition. Al-Risalah was influenced by Pascal, C++, and Eiffel languages and is intended to teach Arabic-speaking students how to program and how to understand the concepts of object-oriented programming. Lately, couple of other Arabic programming languages were developed including AMORIA [21], Ebda3 [22], Jeem [23], Loughaty [24], Qlb [25], and Kalimat [26].

Unfortunately, all the aforementioned Arabic programming languages are not fully comprehensive in that some stayed on papers, others are not turning complete, while others are not compiled. Furthermore, some of these languages are not general-purpose and lack many elementary programming features. Also others are console-based missing graphical user interface features and event handling. Finally, last but not least, the majority of those languages are non-distributable in that they don't generate standalone executable files for Windows or for any other target operating system.

### III. PHOENIX – THE PROPOSED ARABIC PROGRAMMING LANGUAGE

Phoenix is a General-Purpose, High-Level, Imperative, Object-Oriented, Compiled, Arabic computer programming language intended to write computer programs in the Arabic language. Phoenix is C# syntax-like language using modern programming features to improve the programming experience in the Arabic language. Phoenix is compiled in that it generates an object/machine code from source-code prior to program execution. Phoenix in its current implementation runs over Windows operating system and is able to generate an executable file from compiled machine code. Moreover, Phoenix is powered by an easy-to-use and ergonomic IDE (Integrated Development Environment) that allows programmers to create, save, debug, and compile their source-code.

### IV. THE LANGUAGE FEATURES

Phoenix supports many modern and powerful programming features and disciplines making it suitable in the world of software development. They can be summarized as follows:
• Strong data types: Decimal and String
• Implicit type conversion between data types
• Dynamic arrays with automatic bound checking
• Global and Local variable declaration
• Conditional Structures (if and if-else)
• Control Structures (while)
• Code blocks and Compound statements
• Global, local, and function scopes
• Function declaration with parameters and return type
• Recursion
• Arithmetic calculation: + , - , * , / , % , ( )
• String concatenation
• Logical evaluation using && and || operators
• Single line code comments
• Classes, objects, encapsulation
• Access modifiers public, private
• Composition
• Automatic Garbage Collection
• Graphical forms with input and output dialogs

## V. THE COMPILER

The Phoenix compiler consists of six building blocks: Preprocessor, Scanner, Parser, Semantic Analyzer, Code Generator, and Linker [27].

- *The Preprocessor:* Its purpose is to reduce the complexity of the source-code and make the job easier on the scanner. The preprocessor has many tasks including removing code comments, getting rid of extra white-lines and white-spaces, integrating external libraries, and deleting unused variables.
- *The Scanner:* Its purpose is to tokenize the source-code and divide it into meaningful tokens such as keywords, operators, and data values. The algorithm of the scanner is built upon Finite-State Machine (DFA) [28] and Regular Expressions. The scanner has also access to a Symbol Table implemented as a Linked List data structure. Its purpose is to store variable names and their data types, function names, information about scope, and compiler generated temporaries.
- *The Parser:* Its purpose is to detect syntax errors by performing Syntax Checking against the tokens generated by the scanner. The output is a Parse-Tree known as Syntax-Tree. Syntax Checking is about verifying that the arrangement of tokens as received from the scanner are in the correct order and comply with the grammar of the programming language. The algorithm of the parser is a Top-Down Parsing using Recursive Descent Traversal with early error detection.
- *The Semantic Analyzer:* Its purpose is to perform Semantic Checking which consists of verifying that the written source-code comply with the semantics of the programming language. Semantics are the different rules that define restrictions on the syntax. For instance, one of the semantics restricts the use of variables before being declared. Likewise, another semantics restricts calling a function with the wrong number of parameters.
- *The Code Generator:* Its purpose is to convert the parse-tree generated by the parser into a target code. The target code can be either Assembly code, Machine code, Bytes code, or even another high-level language. The current implementation of Phoenix produces Machine code compatible with x86/x64 instruction set architecture.
- *The Linker:* Its purpose is to convert the target code into a native executable code compatible with the underlying operating system. The current implementation of Phoenix generates ".exe" standalone applications compatible with Microsoft Windows.

### A. The Scanner DFA

The algorithm of the scanner is built based on Deterministic Finite Automaton (DFA) and a set of Regular Expressions. Figure 1, 2, and 3 are sample Finite Automata that the scanner uses to detect and tokenize an identifier/variable names, numeric values, and string values respectively.

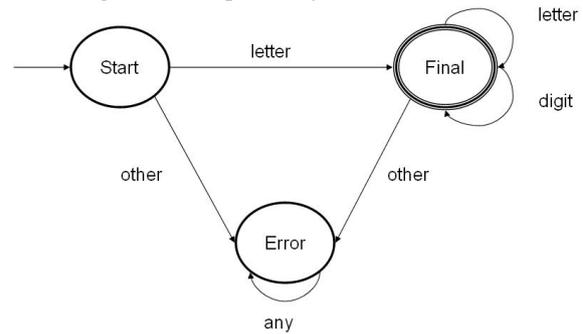

**Fig 1: Finite Automata for Identifiers**

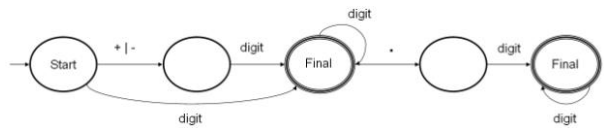

**Fig 2: Finite Automata for Numeric Values**

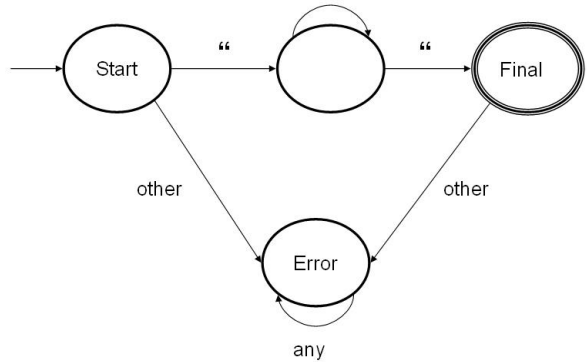

**Fig 3: Finite Automata for String Values**

The language Keywords are also detected by the scanner. They are listed below:

، رقم ، كلمة ، قائمة‑رقم ، قائمة‑كلمة ، وظيفة ، نهاية الوظيفة ،
، صنف ، عام ، خاص ، إذا ، أما عدا ذلك ، كرّر ، أعرض ، أدخل
إستدعاء ، عودة

### B. The Parser Context-Free Grammar

The parser is built upon a formal grammar. The Phoenix parser is based on a CFG or Context-Free Grammar [29] as it provides powerful features including but not limited to recursion, cascading, and nesting. Below is the CFG of the Phoenix parser:

program → function-decl | declaration-stmp | declaration-class

function-decl → وظيفة : ID
( return-type , parameter-list ) {
statement-list } نهاية الوظيفة

return-type → رقم | كلمة | قائمة‑رقم | قائمة‑كلمة

parameter-list → type ID

statement-list → statement-list statement | statement
statement → declaration-stmp
| assignment-stmp
| comparison-stmp
| repetition-stmp
| outputDialog-stmp

| inputDialog-stmp

declaration-class → صنف **ID**
{ access-mod declaration-stmp |
access-mod function-decl }

access-mod → خاص | عام

declaration-stmp → var-declaration | object-declaration | array-declaration ;
var-declaration → type **ID** = **value**
array-declaration → type **ID[NUM]** = { value-list }
object-declaration → **ID ID**
value-list → **NUM ,** value-list | **NUM**
value-list → **STRING ,** value-list | **STRING**
type → رقم | كلمة | قائمة-رقم | قائمة-كلمة

assignment-stmp → assignmentNum-stmp | assignmentString-stmp;

assignmentNum-stmp → var = expression
var → **ID** | **ID** [expression]
expression → ( expression ) addop term | term
expression → ( expression ) mulop term | term
addop → + | -
mulop → × | ÷ | %
term → **NUM | ID | ID** [expression]

assignmentString-stmp → var = expressionString
var → **ID** | **ID** [expression]
expressionString → expressionString concatop term | term
concatop → **&**
term → **STRING | $ | ID | ID** [expression]

comparison-stmp → إذا : comp-expression statement
| أما عدا ذلك إذا : comp-expression statement
statement
comp-expression → expression relop expression |
expressionString relop-str expressionString
relop → == | != | > | < | <= | >=
relop-str → == | !=

repetiton-stmp → كرّر : comp-expression statement
comp-expression → expression relop expression
relop → == | != | > | < | <= | >=

outputDialog-stmp → أعرض : expressionString ;
inputDialog-stmp → أدخل : var ، **STRING**
var → **ID** | **ID** [expression]

**ID** = letter (digit | letter)*
**NUM** = = ((+ | -) digit | digit) digit* . digit digit*
**STRING** = " letter* "
**letter** = أ | ب | .. | ي
**digit** = 0 | .. | 9

## VI. EXPERIMENTS & SAMPLE PROGRAM

In this section, we will write the first computer program using the Phoenix Arabic programming language. It is a sample code that computes the average of a set of five grades or numbers while illustrating the use of function calls, variables, while loop, arithmetic operations, and display dialogs. Figure 4 shows the source-code of the sample program written using Phoenix; while Figure 5 shows its equivalent code written using C#.NET. Finally, Figure 6 is a screenshot for the Integrated Development Environment (IDE) that is used to write, edit, and compile source-code using the Phoenix programming language.

```
وظيفة معدل (-) : البداية
{
    رقم علامة = 0 ؛
    رقم مجموع = 0 ؛

    رقم عداد = 0 ؛
    كرّر : عداد > 5
    {
        أدخل : علامة ، "أدخل علامتك" ؛
        مجموع = مجموع + علامة ؛
        عداد = عداد + 1 ؛
    }

    أعرض : "المعدّل هو " & (مجموع÷عداد) ؛
}
نهاية الوظيفة
```

**Fig 4: Source-Code written using Phoenix**

```
void average()
{
    double grade = 0;
    double total = 0;

    int counter = 0;
    while (counter < 5)
    {
        grade = Convert.ToDouble(Interaction.InputBox("Enter Grade"));
        total = total + grade;

        counter++;
    }

    MessageBox.Show("Average is: " + (total / counter));
}
```

**Fig 5: Equivalent Source-Code written using C#**

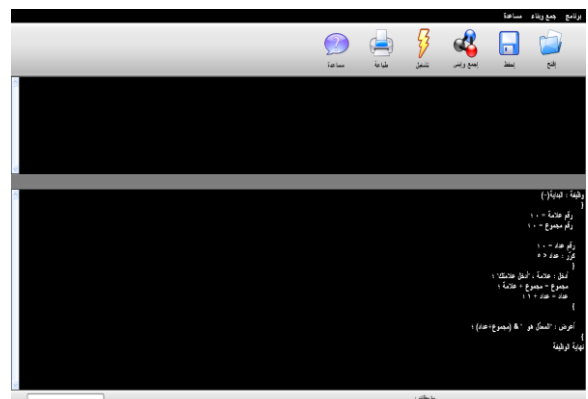

**Fig 6: IDE for Phoenix**

## VII. CONCLUSIONS

This paper discussed the design of a new programming language called Phoenix. It is a General-Purpose, High-Level, Imperative, Object-Oriented, Compiled, and Arabic computer programming language. Phoenix is C# syntax-like and is supported by an Integrated Development Environment. The core of Phoenix is a compiler system made up of six building blocks including a Preprocessor, a scanner based on DFAs and regular expressions, a parser based on a context-free grammar, a semantic analyzer, a code generator, and a linker. The experiments showed a sample program written using Phoenix and its equivalent code written using C#. The results have demonstrated the several powerful features of Phoenix including functions, while-loop, and arithmetic operations and its capability in building general-purpose programs that can be used for real world applications.

## VIII. FUTURE WORK

As future work more advanced object-oriented features are to be investigated such as inheritance, polymorphism, and templates. Moreover, a library of built-in classes and reusable functions is to be developed whose purpose is to provide such capabilities as file processing, database access, graphical user interface, and networking.

## ACKNOWLEDGMENT

This research was funded by the Lebanese Association for Computational Sciences (LACSC), Beirut, Lebanon, under the "Arabic Programming Language Research Project – APLRP2019".